\documentclass[aps,pre,reprint,bibnotes,footinbib,showkeys,groupedaddress,superscriptaddress]{revtex4-1}
\usepackage{graphicx,amsfonts}
\usepackage{amsmath,empheq}
\usepackage{verbatim}
\usepackage{amssymb}
\usepackage{amsthm}

\usepackage{natbib}
\usepackage{epic}
\usepackage{natbib}
\usepackage{xcolor}
\usepackage{color}
\definecolor{darkblue}{rgb}{0,0,0.5}
\definecolor{darkred}{rgb}{0.5,0,0}
\usepackage[colorlinks=true,urlcolor=darkblue,citecolor=darkblue,linkcolor=darkred,hyperfootnotes=false]{hyperref}
\usepackage{textcomp}
\usepackage{cleveref}
\usepackage[normalem]{ulem}
\newcommand{\M}[1]{\mathcal{#1}}

\newcommand{\req}{\rho_{eq}}
\newcommand{\vreq}{\varrho_{eq}}
\newcommand{\bessk}{\text{K}}

\begin{document}

\title{Heat leakage in equilibrium processes}

\author{Reinaldo  Garc\'{i}a-Garc\'{i}a}
\email{reinaldomeister@gmail.com}
\affiliation{Universit\'e Grenoble Alpes, CNRS, ISTerre, 38000 Grenoble, France}
\affiliation{PMMH, CNRS UMR 7636, PSL Research University, ESPCI, 10 rue de Vauquelin, 75231 Paris cedex 05, France}

\begin{abstract}	
The difference between the zero-mass limit of the heat exchanged with a thermal reservoir, and its value as determined from overdamped dynamics,
is termed `heat leakage' or `hidden heat' in the Smoluchowski limit. If present, heat leakages are the sign of the unsuitability of the overdamped approximation
for addressing thermodynamics.
It is accepted that no hidden heat arises in an isothermal process driven by conservative forces. Here, we challenge that conclusion.
The heat exchanged with a reservoir in any isothermal and quasistatic process
connecting two equilibrium states, indeed exhibits hidden contributions. Our results imply that the
overdamped dynamics misrepresents thermodynamics quite generally.
Surprisingly, the hidden heat is described
by an universal distribution in slow processes, easing the correction of the heat statistics in that context.
\end{abstract}

\maketitle

Coarse-graining procedures based on the elimination of fast variables, represent a powerful tool to reduce the inherent complexity of many statistical models. However, hidden variables
may still play a role in coarse-grained systems~\cite{Zamponi,1742-5468-2007-09-P09012,doi:10.1063/1.2907242,PhysRevE.85.041125,
PhysRevLett.108.220601,PhysRevE.91.012130,PhysRevE.91.052128,PhysRevE.93.032103}. It is for that reason that assesing the thermodynamic consistency of such reduced descriptions 
constitutes a genuine concern. For instance, Hondou and Sekimoto found already several years ago that an overdamped Brownian engine 
operating in presence of a spatially varying temperature profile could never attain Carnot efficiency due to an irreversible heat flow
linked to the momentum variable, which persists even in the overdamped limit~\cite{Sekimoto-Carnot}.
Such a situation is surprising at a first sight, because dynamic quantities (trajectories) are in general well represented by the overdamped approximation.

The interest in this type of anomaly 
has recently resurfaced in connection with passive systems out of equilibrium~\cite{Celani-anomaly, MurashitaEsposito, PhysRevE.97.022131,
PhysRevE.98.022102, 2018arXiv180509080P}, and active particles (see for instance~\cite{shankar2018hidden} and~\cite{2018arXiv180800247G}).
The common feature of all these studies is the persistence of irreversible momentum relaxation induced either by spatially varying temperature profiles,
by switching among several thermal reservoirs, or by some internal self-propulsion mechanism. It is now clear that in such strong nonequilibrium settings, the overdamped
approximation misrepresents thermodynamics.

In support of the role of irreversibility, it has been recently concluded that the heat exchanged between a harmonic oscillator and a thermal reservoir during an isothermal transformation
is well captured by the overdamped dynamics, explicitly linking the emergence of heat leakages, for instance, to processess with changing bath temperature~\cite{PhysRevE.97.022131}.
In this Rapid Communication, we show that those conclusions need to be taken carefully. More precisely, we find that the heat exchanged with a reservoir in an isothermal and quasistatic process, exhibits hidden contributions.
With our choice of conditions, we explicitly eliminate all sources of non-equilibrium behavior to
engineer an scenario where such leakages are in principle not to be expected, yet they are present and the overdamped dynamics fails to properly account for heat fluctuations. 

Let us summarize our findings.
In a quasistatic isothermal process connecting an intial equilibrium state $A$ and a final equilibrium state $B$, the zero-mass limit of the
heat distribution computed from the underdamped dynamics, $P_{A\to B}^{(0^+)}(Q)$, and the heat distribution computed directly from the overdamped dynamics, $P_{A\to B}^{(0)}(Q)$, are different. 
This corresponds to the emergence of hidden heat fluctuations even in absence of explicit irreversibility. The two distributions are related via the convolution formula:
\begin{equation}
 \label{main}
 P_{A\to B}^{(0^+)}(Q)=\int_{-\infty}^{\infty}\,\beta\Omega_{Nd}\big(\beta Q_h\big)P_{A\to B}^{(0)}(Q-Q_h)\,dQ_h,
\end{equation}
where $\beta=1/T$ and $T$ is the fixed temperature at wich the process occurs (Boltzmann constant is set
to $1$ throughout this Rapid Communication); $N$ is the number of particles, and $d$ is space dimensionality.
The function $\Omega_\nu(z)$, with $\nu=Nd$, is normalized to $1$, is independent of the path from $A$ to $B$ as far as it is quasistatic, and is also independent of the identity of the end states, $A$ and $B$.
It represents the probability density function of the hidden heat and it is universal in the sense that it is only parametrized by
the number of degrees of freedom and does not depend on details such as the shape of 
inter-particle interactions. $\Omega_\nu(z)$ is expressed in terms of the modified Bessel function of the second kind as
\begin{equation}
 \label{main-a}
 \Omega_\nu(z)=\frac{1}{\sqrt{\pi}\Gamma\big(\nu/2\big)}\bigg(\frac{|z|}{2}\bigg)^{\frac{\nu-1}{2}}\bessk_{\frac{\nu-1}{2}}(|z|).
\end{equation}

Some additional comments are in order. First, one can observe that the hidden heat has zero mean in quasistatic isothermal processes,
implying that any estimation concerning only mean values, remains correct. This helps to understand the conclusions drawn in Ref.~\cite{PhysRevE.97.022131}, 
where the authors only focused on the mean value of the heat.
As a second important remark, we will see that the work distribution is anomaly-free and is accurately captured by the overdamped approximation, in accordance with the results of 
Ref.~\cite{2018arXiv180509080P}, where the same conclusion was drawn from the study of more generic driving protocols in presence of non-conservative forces.

To gain some intuition, we start by analyzing the simplest equilibrium system one can think of.
The discussion below is very illustrative because instead of considering a process, we will only look at the heat fluctuations without even perturbing the equilibrium state of the system. Yet, we find
that the overdamped dynamics does not capture \emph{almost any} of the features of the heat fluctuations in this case.
It is known that the net heat flow between a system and a reservoir vanishes in equilibrium. This result is so far well reproduced,
however, the analysis of the second moment of the heat already reveals a strong discrepancy between overdamped and underdamped descriptions.
We remark that the study of the fluctuations of any thermodynamic quantity beyond its mean value is at the heart of 
stochastic thermodynamics~\cite{Seifert-Review, *SEIFERT2018176} and, correspondingly, is meaningful here. 

\begin{figure}[t]
  \includegraphics[scale=0.3]{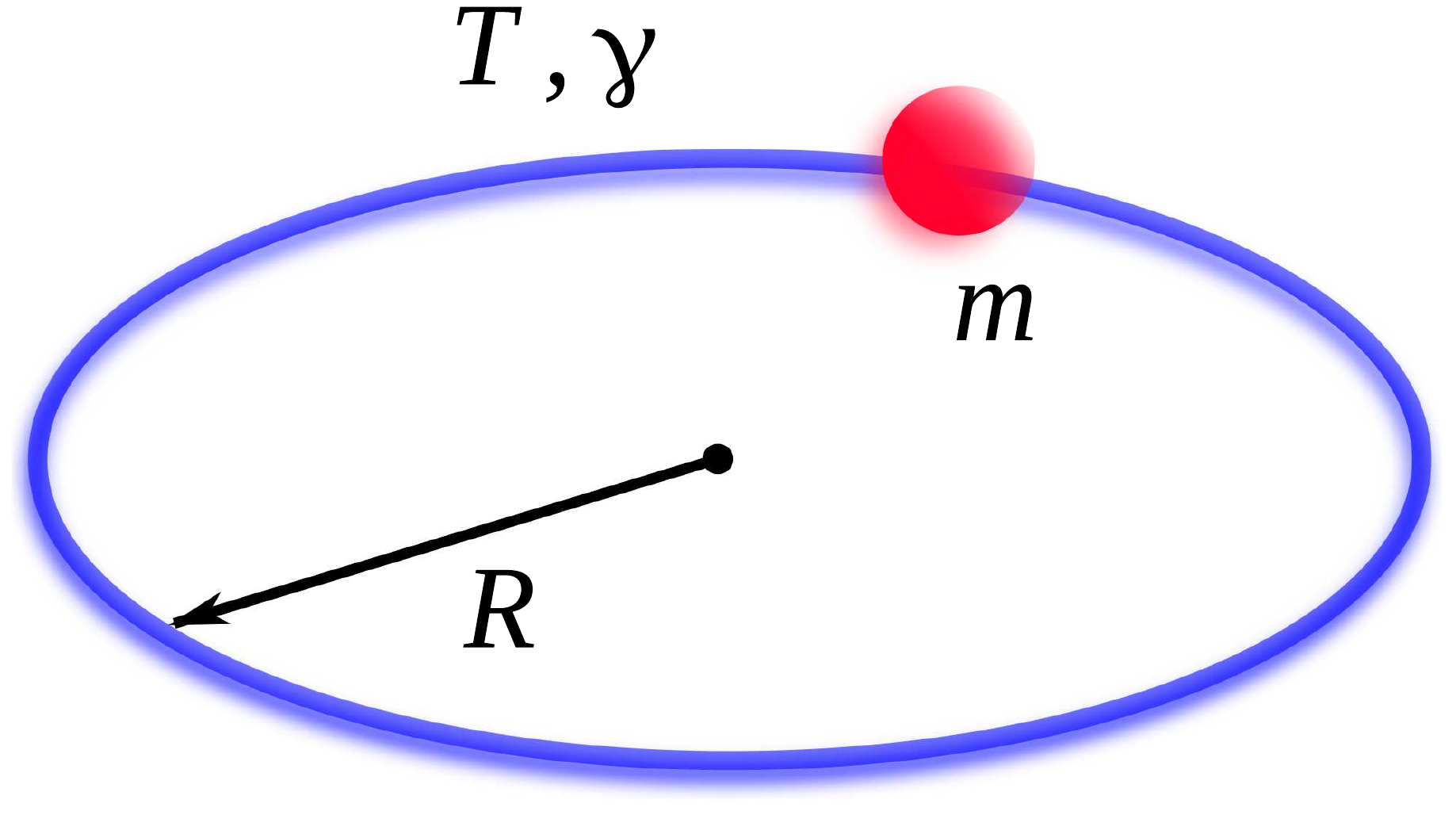}
  \caption{Free Brownian particle of mass $m$ moving on a circle of lenght $L=2\pi R$ immersed in a medium in equilibrium at temperature $T$.}
  \label{fig1}
\end{figure}

Consider a free Brownian particle diffusing on a circle and
immersed in a fluid at temperature $T$ (see Fig.~\ref{fig1}).
In presence of an external bounded potential, the steady state of such system is generically out of equilibrium
because Boltzmann distribution does not comply with the periodic boundary conditions. Nevertheless, the free particle
case exhibits a \emph{bona fide} equilibrium measure.
The Hamiltonian of the free particle consists only of its kinetic energy, $\M K=p^2/2m$, where $m$ is the mass of the particle. 
The associated underdamped stochastic dynamics are then given by $\dot x=p/m$, $\dot p=-(\gamma/m) p+\sqrt{2\gamma T}\xi$, where $\gamma$ is the viscous
friction coefficient of the particle in the medium and $\xi$ is a Gaussian thermal
noise of zero mean and variance $\langle\xi(t)\xi(t')\rangle=\delta(t-t')$.
On the other hand, the overdamped dynamics, 
which formally arises as the $m \to0^+$ limit (Smoluchowski limit) of the underdamped evolution, is simply given by $\dot x=\sqrt{2\mu T}\xi$, where $\mu=1/\gamma$ is the mobility of the particle.

We now rely on Sekimoto's stochastic energetics formalism~\cite{sekimoto2010stochastic}. Consider a time interval, $t\in[0,\tau]$ 
and multiply the momentum evolution equation by $p/m$. Integrating (in the Stratonovich sense) in the given time interval, one then obtains the stochastic version of the first
law of thermodynamics, $\Delta\M K(\tau)=-\M Q_m(\tau)$, where $\Delta\M K(\tau)=(p^2(\tau)-p^2(0))/2m$ is the kinetic energy variation, while
$\M Q_m(\tau)=\int_0^{\tau}(p/m)\circ(\gamma p/m-\sqrt{2\gamma T}\xi)dt$ is the stochastic heat exchanged with the reservoir (with $\circ$ denoting Stratonovich product). Similar analysis of the overdamped dynamics yields
$\M Q_0(\tau)\equiv0\;\forall\tau$ in that case. 

We use the subindex $m$ ($0$) to indicate that the heat is calculated in the underdamped (overdamped) system with mass $m$ (with zero mass).
If the initial condition is sampled from the equilibrium distribution, one has $\langle\M Q_\alpha(\tau)\rangle=0$ for $\alpha=0,m$, as expected. Additionally, one also has
$\langle\M Q_0(\tau)^2\rangle=0$, because the exchanged heat is identically zero in the overdamped case. However, one has $\langle\M Q_m(\tau)^2\rangle\neq0$. 
Indeed, as follows from the underdamped dynamics, the linear momentum is described by an Ornstein-Uhlenbeck process, 
with equilibrium distribution $f_{eq}(p)=(2\pi m T)^{-1/2}\exp(-p^2/2mT)$, and propagator 
$$
g(p,t|p_0)=\frac{1}{\sqrt{2\pi m T\,\Sigma_m(t)}}\exp\bigg(-\frac{(p-p_0\theta_m(t))^2}{2 m T\,\Sigma_m(t)}\bigg),
$$
where $\theta_m(t)=\exp(-\gamma t/m)$, and $\Sigma_m(t)=1-\theta_m^2(t)$. We then rely on the first law of thermodynamics, to compute
$\langle\M Q_m(\tau)^2\rangle=(4m^2)^{-1}\langle(p(\tau)^2-p(0)^2)^2\rangle=(4m^2)^{-1}\int\int\,(p^2-p_0^2)^2g(p,\tau|p_0)f_{eq}(p_0)dp_0\,dp$. The double Gaussian integral
can be readily performed, to give $\langle\M Q_m(\tau)^2\rangle=T^2\Sigma_m(\tau)$. If we now note that $\lim_{m\to0^+}\Sigma_m(\tau)=1\;\forall \tau>0$, and denote
$\langle\M Q_{0^+}(\tau)^2\rangle=\lim_{m\to0^+}\langle\M Q_m(\tau)^2\rangle$,
we have
$
 \langle\M Q_{0^+}(\tau)^2\rangle=T^2\neq0\equiv\langle\M Q_0(\tau)^2\rangle,
$
illustrating that, indeed, the heat fluctuations are misrepresented in the overdamped approximation even in such a simple, non-driven system.

We now proceed with the derivation of our main results. Consider a system of $N$ interacting particles in a medium in equilibrium at temperature $T$. Spatial dimensionality is arbitrary and denoted by $d$, 
and the system is controlled by a set of external parameters that we denote as $\ell$. 
When these parameters are held fixed, the underlying dynamics satisfies detailed balance and the system relaxes to an unambiguously determined equilibrium state. In other words, each equilibrium state of $N$
particles at temperature $T$ is fully parametrized by some choice of $\ell$.
Let $X=\{x_i\}_{i=1}^N$ denote the ensemble of all positions and $\Pi=\{p_i\}_{i=1}^N$ the set of all momenta, with $x_i\in\mathbb{R}^d$ and $p_i\in\mathbb{R}^d\;\forall i$. 
The system Hamiltonian is $\M H(X,\Pi;\ell)=\sum_{i} p_i^2/2m_i+V(X;\ell)$, where $V$ includes all the external potentials and the interactions among the particles. The corresponding stochastic dynamics are
\begin{align}
 \label{under-dyn}
 \dot x_i &=\frac{p_i}{m_i},\nonumber\\
 \dot p_i &=-\frac{\gamma_i}{m_i}p_i-\partial_{x_i}V(X;\ell)+\sqrt{2\gamma_i T}\xi_i.
\end{align}

Here, $m_i$ and $\gamma_i$ are respectively the mass and the viscous friction coefficient associated to the $i$-th particle, and $\{\xi_k\}_{k=1}^N$ are thermal noises with
zero mean and variance $\langle\xi_{k\sigma}(t)\xi_{l\sigma'}(t')\rangle=\delta_{kl}\delta_{\sigma\sigma'}\delta(t-t')$, with $\sigma,\sigma'=1,2,\ldots,d$.
The form of potential energy $V(X;\ell)$ can be assumed arbitrary
as far as compa\-tibility with normalization of the equilibrium distribution, $\req(X,\Pi;\ell)\propto\exp[-\beta\M H(X,\Pi;\ell)]$,
is guaranteed.
In addition, we also consider the overdamped dynamics
\begin{equation}
\label{over-dyn}
 \dot x_i=-\mu_i\partial_{x_i}V(X;\ell)+\sqrt{2\mu_i T }\xi_i,
\end{equation}
with $\mu_i=1/\gamma_i$. The equilibrium distribution in the overdamped case
depends only on the spatial degrees of freedom, $\vreq(X;\ell)\propto\exp[-\beta V(X;\ell)]$ (note that the inertial relaxation times,
$\tau_i=m_i/\gamma_i$, va\-nish for $m_i\to0^+$, and momenta are eliminated as fast variables).

The stochastic heat exchanged with the reservoir can be computed using the same procedure as in the example above. Consider an arbitrary process $\ell(t)$ in the time interval $t\in[0,\tau]$. One has
in the underdamped case $\Delta\M H(\tau)=\M W(\tau)-\M Q_m(\tau)$, where $\Delta\M H(\tau)=\M H(X(\tau),\Pi(\tau);\ell(\tau))-\M H(X(0),\Pi(0);\ell(0))$ is the change of the total energy of the system,
$\M W(\tau)=\int_0^\tau\,\dot \ell\,\partial_\ell V(X;\ell)dt$ is the Jarzynski work~\cite{Jarzynski-a,*Jarzynski-b}, and
$\M Q_m(\tau)=\sum_i\int_0^{\tau}(p_i/m_i)\circ(\gamma_ip_i/m_i-\sqrt{2\gamma_i T}\xi_i)dt$ is the  heat. For the overdamped dynamics one has
$\Delta V(\tau)=\M W(\tau)-\M Q_0(\tau)$, where the work is given by the same expression above, and the heat reads $\M Q_0(\tau)=\sum_i\int_0^{\tau}\dot x_i\circ(\gamma_i\dot x_i-\sqrt{2\gamma_i T}\xi_i)dt$. 
$\Delta V(\tau)=V(X(\tau);\ell(\tau))-V(X(0);\ell(0))$ accounts for the variation of the potential energy only, which is the total energy associated to the overdamped dynamics.

Consider now two equilibrium states $A$ and $B$ respectively parametrized by $\ell_A$ and $\ell_B$. We focus on quasistatic processes starting in $A$ and ending in $B$. One can picture each of such processes
as a number $M\to\infty$ of jumps $\{\delta\ell_n\}_{n=1}^M$, with $\delta\ell_n\to0\;\forall n$. The time between two consecutive jumps is sufficiently long and during that lapse, the parameters 
are held constant, ensuring equilibration. Then the family of intermediate equilibrium states visited along a given quasistatic process is determined by the associated sequence. 
Introduce the joint probability distribution of the work and the 
microstate of the system at the end of the process, conditioned on the initial microstate. In the underdamped system, we denote that quantity by
$R_{A\to B}^{(m)}(X,\Pi,W|X_0,\Pi_0)$, where the initial microstate is sampled from the initial equilibrium distribution, $\req(X_0,\Pi_0;\ell_A)$. For the overdamped dynamics, we denote
the related conditional probability distribution by $R_{A\to B}^{(0)}(X,W|X_0,)$, with the initial positions sampled from $\vreq(X_0;\ell_A)$. Clearly, in the overdamped case the microstates are determined by the positions
of the particles only. We then have the following identity 
(see details in~\footnote{See Supplemental Material at [URL will be inserted by the publisher]})
\begin{equation}
 \label{important}
 R_{A\to B}^{(m)}(X,\Pi,W|X_0,\Pi_0)=\Phi_{eq}(\Pi)R_{A\to B}^{(0)}(X,W|X_0),
\end{equation}
where $\Phi_{eq}(\Pi)=\exp(-\beta(\sum_ip_i^2/2m_i-\Lambda))$, is the equilibrium Maxwell-Boltzmann distribution corresponding to momenta and evaluated in the final microstate,
with $\Lambda=(d/2\beta)\sum_i\ln(\beta/2\pi m_i)$, so that $\Phi_{eq}$ is normalized. 

After all these necessary preparations, we are in conditions to finally demonstrate
our main results. For simplicity, let us first consider the work distributions. In the underdamped case we can write
$\M P_{A\to B}^{(m)}(W)=\int R_{A\to B}^{(m)}(X,\Pi,W|X_0,\Pi_0)\req(X_0,\Pi_0;\ell_A)dX_0d\Pi_0\,dXd\Pi$,
while for the overdamped case we have
$\M P_{A\to B}^{(0)}(W)=\int R_{A\to B}^{(0)}(X,W|X_0)\vreq(X_0;\ell_A)dX_0\,dX$. 
Using Eq.~\eqref{important} together with these definitons and the factorization property, $\req=\Phi_{eq}\vreq$, we have
\begin{align}
 \label{work-dist}
 \M P_{A\to B}^{(m)}(W) &=\bigg(\int\Phi_{eq}(\Pi)\Phi_{eq}(\Pi_0)d\Pi_0d\Pi\bigg)\M P_{A\to B}^{(0)}(W)\nonumber\\
 &\equiv\M P_{A\to B}^{(0)}(W),
\end{align}
independently of the precise values of the mass of the particles. In particular, this implies that after taking the limit $\{m_i\}\to0^+$, one finds 
that the work distribution is correctly represented by the overdamped dynamics:
\begin{equation}
 \label{work-dist-main}
 \M P_{A\to B}^{(0^+)}(W) = \M P_{A\to B}^{(0)}(W).
\end{equation}

It is important to clarify that Eq.~\eqref{important} is not a general result (correspondingly, neither Eq.~\eqref{work-dist}); they hold here because of 
our focus on quasistatic processes. For more general type of transformations,
the proof of Eq.~\eqref{work-dist-main} cannot rely on~\eqref{work-dist} and more advanced techniques are needed~\cite{2018arXiv180509080P}. The same statement remains true for our analysis
of the heat.
To make sense now of Eqs.~\eqref{main} and~\eqref{main-a}, we introduce the heat distributions for both dynamics. In the underdamped case we write
\begin{align}
 \label{heat-dist-und}
 P_{A\to B}^{(m)}(Q) &=\int\delta\big(Q-W+\Delta\M H\big)\req(X_0,\Pi_0;\ell_A)\times\nonumber\\
 &\times R_{A\to B}^{(m)}(X,\Pi,W|X_0,\Pi_0)dX_0d\Pi_0\,dXd\Pi,
\end{align}
with $\Delta\M H=\sum_ip_i^2/2m_i-\sum_ip_{i0}^2/2m_i+\Delta V$, and $\Delta V=V(X,\ell_B)-V(X_0,\ell_A)$.
In the same spirit, one can write for the overdamped case
\begin{align}
 \label{heat-dist-over}
 P_{A\to B}^{(0)}(Q) &=\int\delta\big(Q-W+\Delta V\big)\vreq(X_0;\ell_A)\times\nonumber\\
 &\times R_{A\to B}^{(0)}(X,W|X_0)dX_0dX.
\end{align}

Note that in writing~\eqref{heat-dist-und} and~\eqref{heat-dist-over}, we made use of the first law of thermodynamics to introduce the heat in terms of the work and the total energy variation along the process.
We can now use again Eq.~\eqref{important} together with~\eqref{heat-dist-und} and~\eqref{heat-dist-over}, and the factorization property of the equilibrium measure. Additionally, 
we recall that the hidden heat accounts
for the mismatch between overdamped and underdamped dynamics, so we also introduce the trivial identity 
$
1=\int \delta(Q_h+\Delta\M K)\,dQ_h,
$
where $\Delta\M K=\sum_ip_i^2/2m_i-\sum_ip_{i0}^2/2m_i$ is the kinetic energy change, which cannot be accessed from the overdamped evolution. Putting all this together, we find
\begin{equation}
 \label{almost-there}
 P_{A\to B}^{(m)}(Q)=\int F(Q_h)P_{A\to B}^{(0)}(Q-Q_h)\,dQ_h,
\end{equation}
where the function $F$ is defined as
\begin{equation}
 \label{F-def}
 F(Q_h)=\int\delta(Q_h+\Delta\M K)\Phi_{eq}(\Pi)\Phi_{eq}(\Pi_0)\,d\Pi_0d\Pi.
\end{equation}

It turns out that a simple rescaling of variables, $p_i=\sqrt{2m_iT}y_i$, $p_{i0}=\sqrt{2m_iT}y_{i0}$, allows to write $F(Q_h)=\beta\M F(\beta Q_h)$, where the function $\M F(z)$ does not depend on the
mass of the particles. Furthermore, one can show that $\M F(z)\equiv\Omega_{Nd}(z)$, with $\Omega_{\nu}$ given by Eq.~\eqref{main-a}; the interested reader can find all the technical details in~\cite{Note1}. This yields
\begin{equation}
 \label{prev-main}
 P_{A\to B}^{(m)}(Q)=\int_{-\infty}^{\infty}\,\beta\Omega_{Nd}\big(\beta Q_h\big)P_{A\to B}^{(0)}(Q-Q_h)\,dQ_h,
\end{equation}
which after taking the limit $\{m_i\}\to0^+$ leads directly to our first main result, Eq.~\eqref{main}.

The procedure followed above and, in particular, Eq.~\eqref{F-def}, are useful to understand why the distribution of the hidden heat is universal in slow processes. Universality arises from the fact
that in quasistatic transformations, the hidden heat is the manifestation of the equilibrium kinetic energy fluctuations not accounted for in the overdamped limit. As inferred from the equipartition
theorem, such fluctuations are only characterized by the temperature and the number of degrees of freedom. All the remaining details of the model are then unimportant, as far as equilibrium
is well defined and dynamically accessible.

We are now in position to address in more detail the apparent contradiction between our findings and those mentioned above from Ref.~\cite{PhysRevE.97.022131}, indicating precisely in which sense
those results have to be applied with care. As mentioned before, the authors of Ref.~\cite{PhysRevE.97.022131} claim that no heat-leakage arises in an isothermal process. 
In this work, we have shown that those claims are correct only in average.
At least in the type of isothermal transformations considered here, there are also hidden heat fluctuations, their particularity is that their mean value vanishes. 
Although we focus on slow transformations, we believe that our conclusions remain valid when considering finite-rate processes too, see the discussion below.
This implies, in particular, that heat distributions are misrepresented by the overdamped approximation \emph{also} in isothermal processes.

Let us add some comments about the generalization of our results beyond quasistatic transformations.
As follows from the first law of thermodynamics, the difference between the work and the heat is simply the energy change in the the process, 
which is a border term in the sense that it does not depend on the full trajectory followed by the system,
but only on the initial and final microstates. 
When a system is driven at a finite rate for a long time, the work behaves as a time-extensive observable, meaning that $W\sim \tau$ in distribution, where
$\tau$ is the duration of the process.
In this context, if the energy variation is bounded, it can be neglected in front of the work, meaning that $W\sim Q$. As shown in this Rapid Communication and
also in Ref.~\cite{2018arXiv180509080P}, the work distribution is always well described by
the overdamped approximation, so, in such asymptotic conditions, the heat too. We believe that this is precisely the scenario in which an equivalence between
underdamped and overdamped dynamics in presence of conservative forces was found in Ref.~\cite{MurashitaEsposito}.
Nevertheless, border terms are in general relevant and cannot be dismissed in many situations, a fact that is
already well known (e.g.~\cite{PhysRevLett.108.240603}). Furthermore, the kinetic energy is only bounded from below, and its variation may be as large as one desires.
For that reason, if the process has a finite duration, kinetic energy fluctuations still play a role and contribute additional heat fluctuations
that cannot be accounted for in the overdamped limit.

In conclusion, we have shown that in isothermal transformations occurring in textbook equilibrium conditions, the heat exchanged with a reservoir is incorrectly determined in the overdamped approximation 
due to additional kinetic
energy fluctuations not taken into account in the Smoluchowski limit. In such slow processes, the hidden heat has zero average and is described by an universal distribution, only parametrized by the number
of degrees of freedom. We have also provided arguments in support of the idea that our results may naturally extend beyond quasistatic transformations, implying that, 
as regards heat fluctuations, the overdamped approximation is never to be trusted.

This work was partially supported by the French National Research Agency (ANR) under grant \textnumero  ANR-16-CE11-0026-03.

\bibliographystyle{apsrev4-1.bst}
\bibliography{./heat.bib}

\end{document}